# Effective structural unit analysis in hexagonal close packed alloys – reconstruction of parent beta microstructures & crystal orientation post processing analysis

Authors

**Ruth Birch[a]* and Thomas Benjamin Britton[ab]**

[a]Materials, Imperial College London, London, SW7 2AZ, United Kingdom

[b]Materials Engineering, University of British Columbia, Vancouver, Canada

Correspondence email: ruth.birch07@imperial.ac.uk

**Funding information** Engineering and Physical Sciences Research Council (award No. EP/S515085/1 to Ruth Birch, Thomas Benjamin Britton); Royal Academy of Engineering (award to Thomas Benjamin Britton).

**Synopsis** We present a method to reconstruct the parent body centred cubic microstructure from a child hexagonal closed packed microstructure of zirconium alloys. This is used to provide post processing of the microstructure to understand structural units in the material.

**Abstract** Materials with an allotropic phase transformation can result in the formation of microstructures where grains have orientation relationships (ORs) determined by the transformation history. These microstructures influence the final material properties. In zirconium alloys, there is a solid-state body centred cubic (BCC) to hexagonal close packed (HCP) phase transformation, where the crystal orientations of the HCP phase can be related to the parent BCC structure via the Burgers orientation relationship (BOR). In the present work, we adapt a reconstruction code developed for steels which uses a Markov Chain clustering algorithm to analyse electron backscatter diffraction (EBSD) maps and apply this to the HCP and BCC BOR. This algorithm is released as open-source code (via github, as ParentBOR). The algorithm enables new post processing of the original and reconstructed data set to analyse the variants of the HCP alpha-phase that are present and understand shared crystal planes and shared lattice directions within each parent beta grain, and we anticipate that this assists in understanding the transformation related deformation properties of the final microstructure. Finally, we compare the ParentBOR code with recently released reconstruction codes implemented in MTEX to reveal differences and similarities in how the microstructure is described.

## 1. Introduction

Burgers reported on the body centred cubic to hexagonal close packed orientation relationship (OR) in zirconium (Burgers, 1934) that had been previously reported by Vogel & Tonn (1931) using X-ray rotation photograph examination (which would now be called a variant of X-ray Laue diffraction) of very large zirconium crystals. In modern materials science, this OR is the so called "Burgers orientation relationship" (BOR). This transformation is important to understand the processing and performance of titanium, zirconium and hafnium alloys, which all have the allotropic phase diagram for high atomic fractions of the base metal in the alloy.

The presence of related microstructural units on performance is well known and reported in the titanium literature, specifically in terms of cold dwell fatigue (Rugg *et al.*, 2007). Regions of common orientation appear due to the BOR and anisotropy within the strain path which leads to regions of (near) common orientation. Reports of these features varies in the literature, broadly called "microtextured regions" (e.g. (Cappola *et al.*, 2021)), "macrozones" (e.g. (Germain *et al.*, 2008)) or the "effective structural unit" (Rugg *et al.*, 2007). Hafnium and zirconium are less well studied in the open literature, but there are limited reasons why similar processing-microstructure-mechanical property linkages will not exist. The only difference is that the loading conditions and safety considerations in their engineering applications will vary.

In practice, the microstructure of the high temperature phase is often difficult to observe directly. Instead, often the BOR can be observed due to orientation relationships between neighbour α -grains. If the two α -grains are from the same parent β-grain, then the α - α grain boundary will have a 'special' OR. This means the parent β-microstructure can be reconstructed from the α -grain structure, provided there is limited subsequent orientation change in the α domain.

To measure the local orientations between grains, electron backscatter diffraction (EBSD) can be used to create 2D (and 3D) microstructure maps where the orientation and phase of each grain is measured at high spatial resolution (for a brief review, see Wilkinson & Britton (2012). This data makes it possible to explore the BOR within these engineering alloys.

In a material that has a solid phase transformation, e.g. when cooling from a (parent) high temperature β phase to a (child) lower temperature α phase, the child phase orientation may be related to the parent grain orientation if an OR is present. For most ORs, the parent phase may have multiple different options for the transformation product orientation, the so-called variants, due to the symmetry relationships and transformation strain & rotation between the two crystal structures. This is complicated further by the symmetry of the parent crystal and the child crystal which can result in ambiguity in calculating the parent grain orientation from a single child grain. There is a similar challenge in predicting the orientation of the child orientation from an initial parent. However, if more than one child orientation is present (e.g. as the parent grain contained more than one nucleation site for the phase transformation), then it is possible to reduce the potential candidate parent phases.

The challenge of determining the parent grain orientation using EBSD is well explored in the literature, and many manual (Humbert et al., 1994, 1995; Humbert & Gey, 2002; Tari et al., 2013), semi-automated (Gey & Humbert, 2003; Germain et al., 2012), and fully automated (Cayron, 2007; Germain et al., 2012, 2019; Glavicic et al., 2003b; Hadke et al., 2016; Davies, 2009; Nyyssönen et al., 2018) grain reconstruction codes have been developed. Groups of similar grains can be identified using the grain boundary misorientation, mean grain orientations or the pixel orientations in EBSD data. Refinements to this process include iterating the OR to take into account the experimental results or using more complex grain grouping methods, for example, using the Markov clustering algorithm (van Dongen, 2000).

Many reconstruction methods start with ideas shared by Humbert et al. (Humbert et al., 1994, 1995; Humbert & Gey, 2002). These methods start by comparing the experimental grain boundary misorientations to the theoretical grain boundary misorientations (between potential α variants) from the OR (Humbert et al., 2015; Germain et al., 2007). The aim is to identify the parent grain boundaries (i.e. those that don't match the theoretical misorientations between variants), usually as part of a grain grouping step.

A limitation of these methods is often that they do not deal well with experimental deviation from the OR, for example deviations caused by internal stresses from thermal or mechanical processing. Cayron et al. (2006) compared the tolerance for the misorientation of the room temperature phase vs the theoretical misorientations for the OR in titanium/zirconium alloys and steels, and found these to be ≈5 ° and ≈2-4 ° respectively. To overcome this issue, an iterative step for the OR has been introduced in some cases, such as work by Nyyssönen et al. (2016, 2018, 2020) in steels.

The grain grouping step is where a lot of methods deviate, with some codes reconstructing without grouping grains first, and others grouping parent to reconstruction. For example, the triplet method (Krishna et al., 2010) uses the groupoid method (Cayron, 2006; Cayron et al., 2006) to identify triplets of grains with a low tolerance angle. These 'nucleus' triplet grains are then grown by adding neighbouring grains that meet a tolerance condition, with this nucleation and growth cycle repeated until no more nuclei can be found. This is a non-iterative and quick method, but it may not group all of the grains if they do not meet the tolerance condition to be grouped together if there are insufficient 'nucleus' triplets in the data set. Alternatively, there is the cluster method (Hadke et al., 2016) which identifies clusters of the child phase which are highly likely to have originated from the same parent grain, then grows these clusters and back-transforms them to the parent phase using the summation of minimum misorientation angle (SMMA) method (Tari et al., 2013). This is a computationally expensive, iterative method, with good accuracy.

Another potential issue is the artificial reunification of grains, particularly in cases where the grains are grouped parent to transformation. This can be addressed in some methods by adjusting tolerance values or coarsening parameters (e.g. for the MCL algorithm (Gomes & Kestens, 2015; Nyyssönen et al., 2018)). Similarly, uncertain grains (either at the grouping stage or to determine the parent orientation) are often an issue, either due to island grains within larger grains, or just small groups of child grains. These are dealt with in in a number of ways, from not being reconstructed (Cayron et al., 2006; Cayron, 2007), to additional subroutines in the code (e.g. (Germain et al., 2012, 2019)) or back-filling from neighbouring grains (Krishna et al., 2010).

For large datasets, there is a tension between accuracy and computational time, so some methods employ statistical analysis to support parent grain reconstruction. These include work by Glavicic et al. (2003a,b) which uses Monte-Carlo analysis and two misorientation minimisation steps and has the advantage of requiring no a priori knowledge of the parent grain boundary locations or variant selection. Also, work by Gomes & Kestens (2015) and Nyyssönen et al. (2016, 2018, 2020) that employs the Markov clustering algorithm to create a network of discrete clusters of similar grains. This method is computationally efficient, has excellent noise tolerance and only one coarsening parameter. Some groups have also worked to combine several of the above methods, for example, Bernier et al. (2014), which brings together OR refinement (Nyyssönen et al., 2016; Miyamoto et al., 2009; Humbert et al., 2011); local pixel-by-pixel analysis (Miyamoto et al., 2009, 2010); and nuclei identification and spreading (Germain et al., 2012).

In this paper, we first describe the algorithm, which is based upon the parent austenite grains reconstruction code by Nyyssönen *et al.* (2016) using MCL, which we have adapted for two phase BCC-HCP materials. In brief, this algorithm reconstructs the microstructures by identifying grain boundary types, then creating a discrete network of clusters of similar grains using a Markov Chain Cluster analysis of the initial microstructure. The BOR is then used with these clusters to reconstruct the parent grain microstructure. We demonstrate the reconstruction code using a simulated dataset. Next, we apply the method to an experimentally obtained data set from zircaloy-4 which has been quenched from the parent β phase. In this data set, we explore the relationship between parent β grain size, and shared crystallographic features with regards to the phase transformations and mechanical properties.

## 2. Symmetry and crystallography of the Burgers orientation relationship (BOR)

The OR for the BCC/HCP phase transformation in zirconium alloys (as reported by Burgers (1934)) is:

$$(0001)_\alpha // \{110\}_\beta \quad \text{and} \quad \langle 11\bar{2}0\rangle_\alpha // \langle 1\bar{1}1\rangle_\beta \qquad [1]$$

### 2.1. BCC to HCP

For any initial β grain, the transformed α orientation can be one of 12 unique variants, as described by the BOR and this is shown in Figure 1.

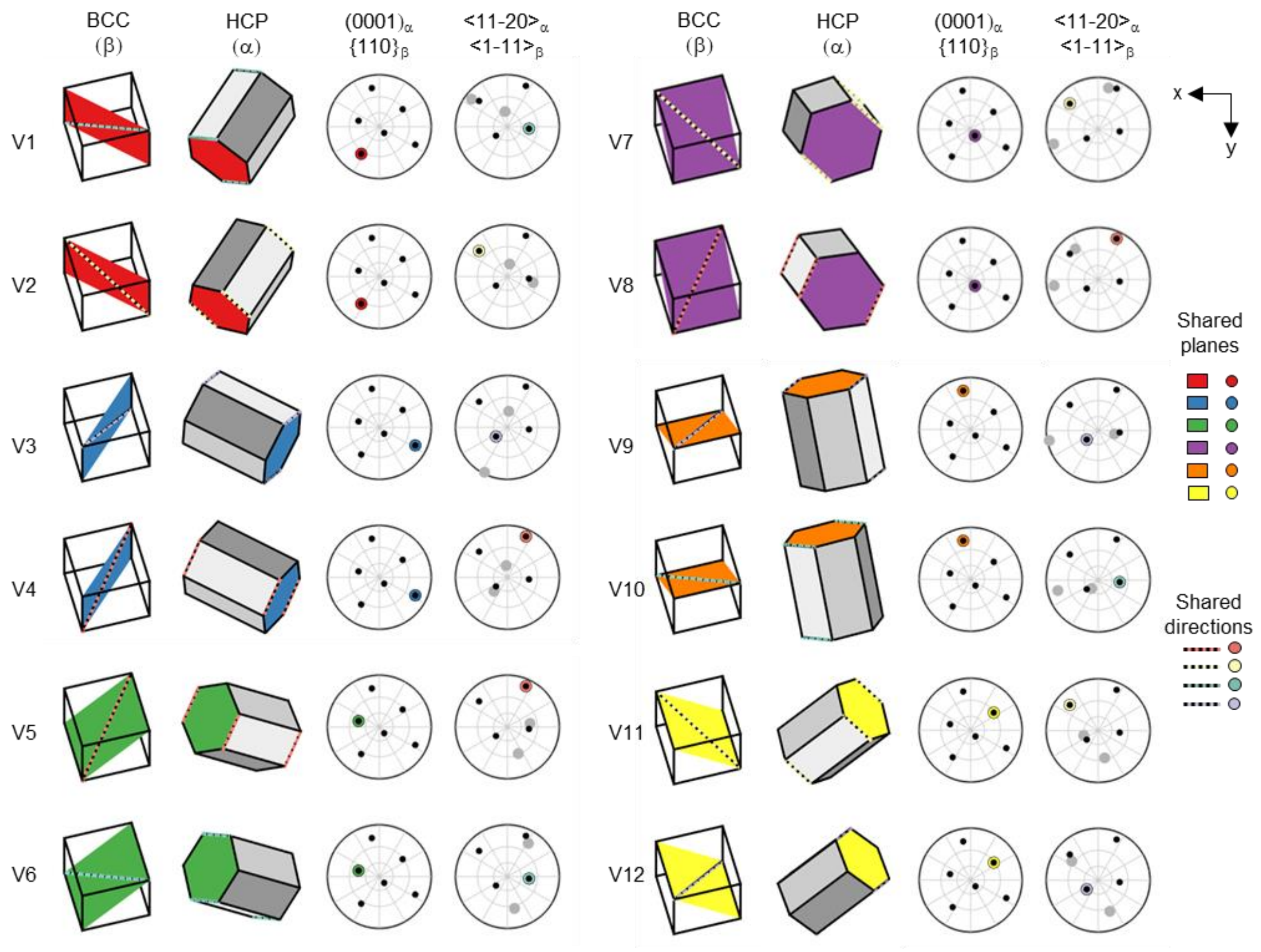

**Figure 1** 12 unique α variants (V1-12) for a fixed β orientation of (0,30,15) showing the corresponding planes and directions between the BCC and HCP orientations via prisms and pole figures. For the prisms: shared planes between variants are indicated by the colour of the plane and shared directions between variants are indicated by the coloured + dashed directions. For the pole figures, the shared plane or direction is shown by the coloured dot, the other HCP points by the larger grey dots and the BCC orientation is shown by the smaller black dots.

In Figure 1, the β phase (BCC) orientation is kept constant, with each of the six $\{110\}_{\beta}$ planes highlighted in turn, in a different colour. Each of these $\{110\}_{\beta}$ planes contain two different $\langle 1\bar{1}1\rangle_{\beta}$ directions, e.g. for $(110)_{\beta}$, the two directions are $[1\bar{1}1]_{\beta}$ and $[\bar{1}11]_{\beta}$. For each cube shown, one of the two $\langle 1\bar{1}1\rangle_{\beta}$ directions contained within has been highlighted with a dashed pastel coloured/black line. The corresponding hexagonal prisms (α orientations - which do change) are shown in the columns adjacent to the cubes. In each case, the $(0001)_{\alpha}$ plane is highlighted with the same colour as the plane that it matches in the BCC orientation and the $[11\bar{2}0]_{\alpha}$ direction has the same pastel coloured/black dashed line as its corresponding $\langle 1\bar{1}1\rangle_{\beta}$ direction in the BCC orientation. This information is shown on the pole figure plots for the BCC/HCP planes and directions. In this case, the BCC orientation (including symmetry) is shown with the small black dots, and the larger dots are the HCP planes & directions. The overlapping, coloured, point is the shared plane or direction and the colours correlate with the prisms. Note that all prisms and pole figures have the same axes.

As can be seen in Figure 1, there are common shared planes and directions between the 12 unique α variants that can result from an initial parent grain orientation. There are four groups of variants that share the same shared direction, as shown by the colour of the directions in the figure, e.g. V1,6 & 10 have a common shared direction. Similarly, there are six pairs of α variants with the same shared plane, e.g. V1 & V2.

### 2.2. HCP to BCC

When the situation is reversed, any start HCP orientation gives rise to 6 potential unique BCC orientations, as illustrated at Figure 2. The HCP crystal is aligned with the $< c >$ axis out of plane and one $< a >$ direction pointing upwards, as per our chosen convention of the reference crystal orientation for HCP. For each $< a >$ direction parallel with a $< 1\bar{1}1 >_\alpha$ in the BCC phase, there are two orientations of the BCC crystal. This can be explained as for each HCP/BCC orientation alignment, such that we align $< a_1 >$ (which is parallel to $< b_1 >$ in the β phase) as a common $-\frac{1}{3}[\bar{1}2\bar{1}0]_\alpha$, then each of the two related variants the $< a_2 >$ and $< b_2 >$ relationship (which is 10.52° misaligned) switches from $-\frac{1}{3}[2\bar{1}\bar{1}0]_\alpha$ to $-\frac{1}{3}[\bar{1}\bar{1}20]_\alpha$. Six variants in total are generated, as this is repeated three times for each of three HCP $< a >$ directions.

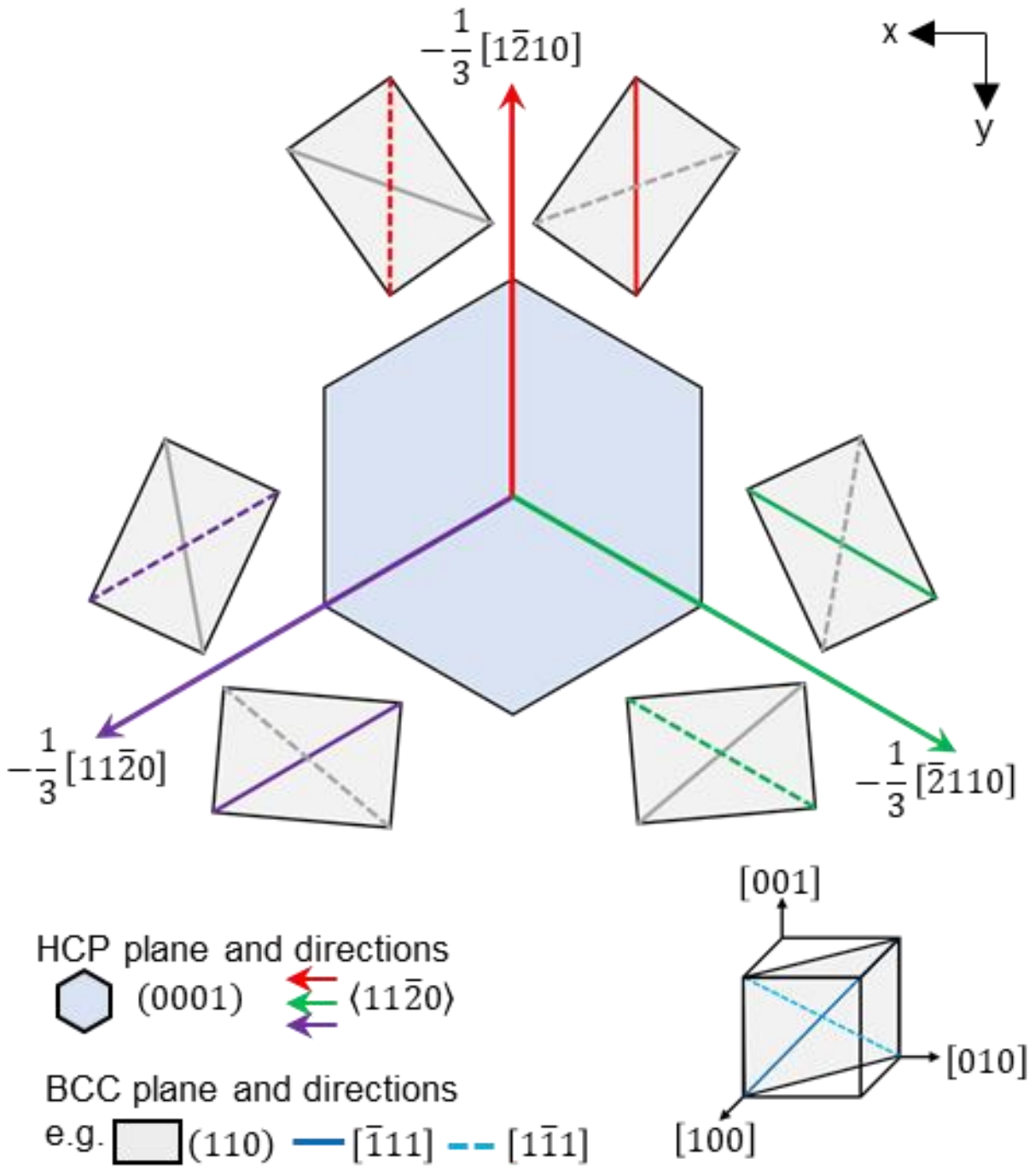

**Figure 2** 6 potential unique parent β orientations for a start α reference orientation. Parallel directions are indicated using the <a> direction colouring.

### 3. Reconstruction algorithm & microstructural post processing

We have developed an MTEX (Bachmann *et al.*, 2010) based code in MATLAB (see the Data Availability Statement). This code can be broken down into two sections: (A) Initial reconstruction and (B) post processing.

### 3.1. Initial reconstruction

This section of the code is adapted from work by Nyyssönen *et al.* (2016, 2018, 2020). The existing algorithm is used to reconstruct the parent austenite microstructure from a child martensite microstructure found in steels. To adapt this code for the BCC to HCP (β→α) phase transformation in HCP alloys, both the phases and OR have been updated (which includes consideration of the reference orientations of the α and β phases). Here we follow the conventions described by Britton *et al.* (2016) where the b axis of the reference HCP unit cell is aligned along the b axis of the reference BCC unit cell. We use the BOR described in Equation 1.

As we have discussed, using this OR, there are 12 potential unique α variants for any start β orientation. In reverse, there are 6 potential unique β variants for any start α orientation (Karthikeyan *et al.*, 2006). This is important in the initial reconstruction and post processing algorithms.

The algorithm for the initial reconstruction section is shown schematically at Figure 3 and contains 4 main steps:

(1) setting up the OR;

(2) identifying grain boundary types;

(3) applying the MCL algorithm;

(4) reconstructing the parent β microstructure.

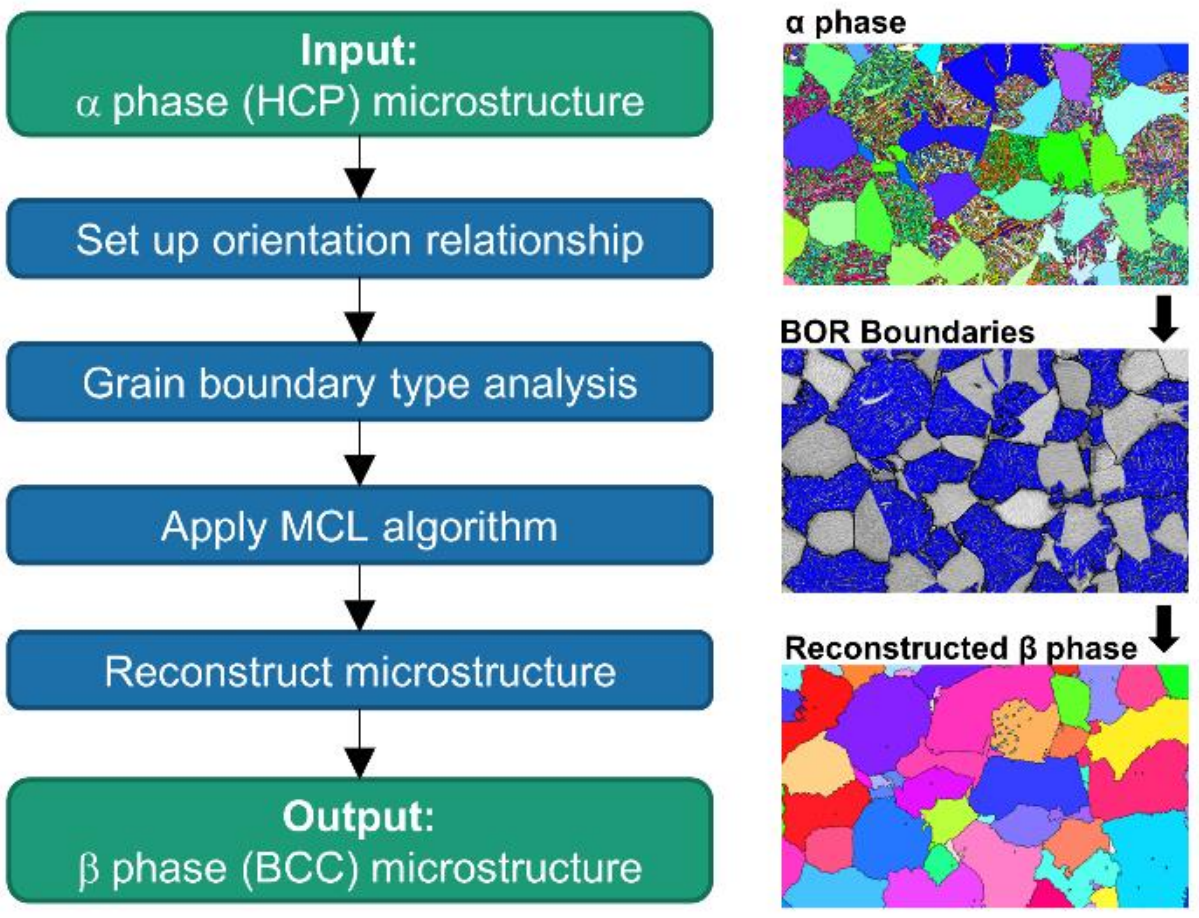

**Figure 3** Initial reconstruction code overview – steps to go from the input α phase microstructure (HCP) to the reconstructed β phase (BCC) microstructure, with microstructure schematics.

### 3.1.1. Setting up the orientation relationship (OR)

Firstly, the OR is established for the β → α transformation. The transformation matrix ($T_{\beta\rightarrow\alpha}$) is set up using an initial pair of orientations ($\alpha_{init}$ (0,45, 24.7) and $\beta_{init}$ (0,0,0)) that meet the BOR. The BOR relationship links the BCC to HCP orientations, and how one of the $< 010 >_\beta$ direction matches onto the HCP unit cell and different EBSD indexing algorithms change these relationships. We follow Britton et al (2016) with $< 1\bar{2}10 >_\alpha$ //Y and $< 010 >_\beta$//Y for the reference orientations[1]..

$$T_{\beta\rightarrow\alpha} = \alpha_{init} \cdot (\beta_{init})^{-1} \qquad [2]$$

This is then expanded to include symmetry:

$$T_{\beta\rightarrow\alpha}^{1-24} = S_\beta^{1-24} \cdot T_{\beta\rightarrow\alpha} \qquad [3]$$

where $S_\beta^{1-24}$ are the 24 symmetry matrices for the BCC phase. This allows the potential α variants ($\alpha^{V1-24}$) to be calculated for any start β orientation (β) using:

[1] Please note that Britton et al. use a convention where <b> is parallel to the Y axis, and in the HCP unit cell <b> is one of the <a> directions.

$$\alpha^{V1-24} = \beta \cdot T^{1-24}_{\beta\rightarrow\alpha} \quad [4]$$

Note that due to symmetry, only the first 12 are unique, the second 12 are repeats.

### 3.1.2. Identifying grain boundary types

The OR is then used to identify parent grain boundaries on the α grain map. A set of α variants is calculated and the misorientations between the variants found. Due to symmetry, the list of misorientations can be minimised to a short list of angle/axis pairs (Gey & Humbert, 2003; Karthikeyan *et al.*, 2006) as detailed at Table 1.

**Table 1** Minimised list of angle/axis pairs for α-zirconium that can exist between α variants from the same parent β grain (as determined from our code, but similar to Gey et al. (Gey & Humbert, 2003), and Karthikeyan et al. (Karthikeyan et al., 2006)).

| ID | Angle (°) | Axis | | | |
|---|---|---|---|---|---|
| | | h | k | i | l |
| 1 | 10.53 | 0 | 0 | 0 | 1 |
| 2 | 60 | -1 | 2 | -1 | 0 |
| 3 | 60.83 | -12 | 7 | 5 | 3 |
| 4 | 63.26 | -2 | 1 | 1 | 1 |
| 5 | 90.00 | -7 | 12 | -5 | 0 |

It is these angle/axis pairs that are used to evaluate whether each grain boundary segment was a parent grain boundary or not. In the experiment, the misorientation across an α-α boundary may not be exact, both due to measurement uncertainty and rotation away from transformation strain. A threshold of 4 ° is set as default. Boundaries which are not described by the special relationships in Table 1 (within the selected threshold) are classified as non-BOR i.e. parent grain boundary.

### 3.1.3. Applying the MCL algorithm

At this point, the Markov clustering algorithm (MCL) (van Dongen, 2000) is applied to identify discrete clusters of closely related α grains that will be used for the reconstruction step. The aim of using a clustering algorithm is to quickly identify the child grains that originated from the same parent grain and speed up the reconstruction process. In brief the MCL enables groups of related BOR α grains to be clustered together for parent β orientation determination. Similar approaches exist (Hadke *et al.*, 2016; Germain *et al.*, 2012), but often require further steps to incorporate grains not included in the initial analysis (due to threshold conditions).

The algorithm creates a network based upon the closeness of grain boundaries to the potential grain boundaries. This network is then put through a series of expansion and inflation steps until a discrete network of clusters is established. For more discussion of the MCL algorithm, please see Nyyssönen *et al.* (2018). An advantage of this algorithm is the single factor for adjusting it – the 'inflation power'. In practice this term can change the number of parent β grains that are found. A smaller inflation power decreases the number of discrete clusters identified, i.e. The user is advised to explore variation of the inflation power and the output reconstruction (which is assisted by the post processor, described later).

#### 3.1.4. Reconstructing the parent β microstructure

Once clusters of α-grains are joined together to describe one parent β grain, it is now valuable to determine the parent β-grain orientations. The reconstruction step combines (1) and (3), with each discrete cluster of α grains treated individually. For each α grain within the cluster, the potential β grain orientations are calculated (using $T_{\beta\rightarrow\alpha}$). When this is completed for every α grain within the cluster, a density function is used to find the most intense orientation – this is allocated as the β orientation for that cluster. Once completed for all discrete clusters, an EBSD map for the parent β phase can be plotted. Our code then processes the grains for the β phase ready for the post processing section.

### 3.2. Post processing

Once the parent β grain boundaries are determined, it is possible to evaluate:

(1) Uniqueness of the determined β orientation.
(2) The shared $< a >_{\alpha}$ directions within the parent β grain.
(3) The shared $\{0001\}_{\alpha}$ planes within the parent β grain.

The additional information generated in this section of the code can then be used to look at the influence of grain boundary types and variant selection.

#### 3.2.1. Certainty of the β-orientation and the associated α variants

The α variants are calculated on a grain by grain (β) basis – this allows for comparison of α variants within each β grain, but not between (β) grains. The process used to analyse the α variants present begins by setting the β grain orientation to the fundamental zone ($\beta_{ori} \rightarrow \beta_{ori}^{FZ}$) and then using this to calculate the potential α variants (${\alpha_{calc}^{V1-12}}$):

$$\alpha_{calc}^{V1-12} = \beta_{ori}^{FZ} \cdot T_{\beta\rightarrow\alpha}^{1-12} \quad [5]$$

The minimum misorientation between each of the calculated α variants and the parent β orientation ($\beta_{ori}^{FZ}$) are then calculated and stored as the 'ideal' misorientations:

$$Miso_{ideal}^{1-12} = \left(\beta_{ori}^{FZ}\right)^{-1} \cdot \alpha_{calc}^{V1-12} \quad [6]$$

Next, the experimental β→α misorientations ($Miso_{exp}^{1-12}$) are calculated for each α grain present ($\alpha_{1-n}$) within the parent β grain ($\beta_{ori}^{FZ}$):

$$Miso_{exp} = \left(\beta_{ori}^{FZ}\right)^{-1} \cdot \alpha_{1-n} \quad [7]$$

Finally, each experimental misorientation is compared with the list of ideal misorientations – the number of the closest match is taken to be the α variant ID for that grain.

#### 3.2.2. β certainty

The number of potential β orientation options (β certainty) for each parent grain identified depends on the number and type (whether there are shared planes or directions) of unique α variants present. For example, if there is only one α grain, it is not possible to determine which of the 6 potential β orientation options is the parent β orientation.
To fully determine the parent β orientation, the number of unique α variants required depends on the type of variants present (shared basal plane, which we call (c) in this paper; or shared <a> direction; or neither). In the case where there is a shared plane, three of the β options will be the same, whilst for a shared direction, two options will be the same between variants. The minimum number of unique α variants required for a unique β-parent is 2 and the maximum is 4 in the case where three of the variants have a common <a> direction. The algorithm used for determining the number of β options is described in Figure 4.

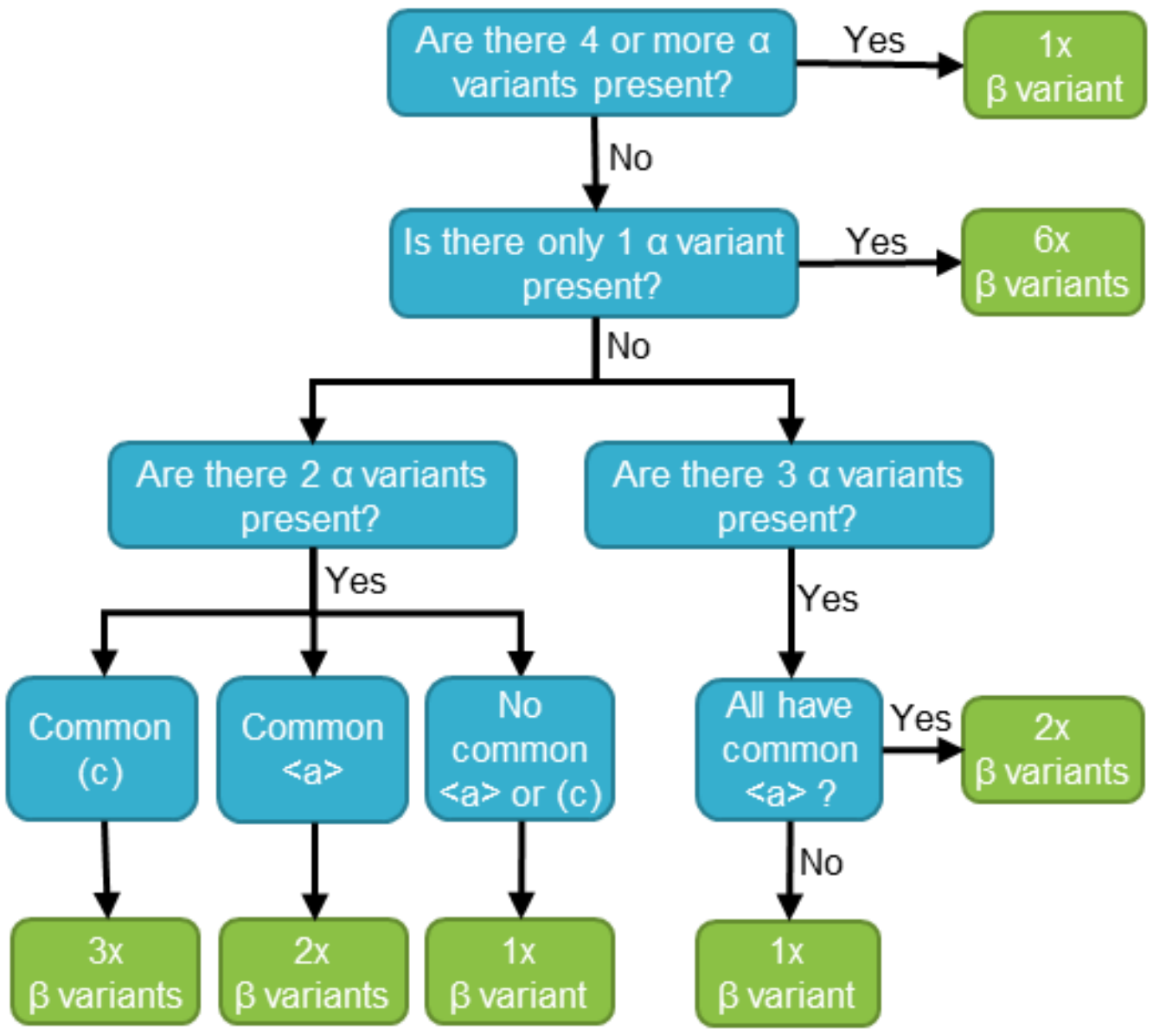

**Figure 4** Decision flowchart for the number of potential unique parent β grain orientations, depending on the number and type of unique α variants present with the parent β grain.

### 3.2.3. β orientation options

As part of the step to calculate the β certainty, all potential parent β orientation options (β options) are stored, as calculated from the associated α grains within the parent β grain. For uniquely determined β grains, this is a single orientation, but where there is uncertainty different orientations can be represented. Additionally, the number of potential β orientations can be reported.

### 3.2.4. Analysis of the structural units within the α-phase: shared planes and directions

With the α variants identified for each parent β grain, variants with shared planes or directions can be identified. Within the set of 12 unique α variants, there are a number of combinations that share a common plane or direction, as illustrated at Figure 1 for a set of α variants calculated from a start β orientation of (0,30,15). This enables the α grain map to be plotted coloured by variant number (coded with respect to each parent β grain), as well as the shared $\{0001\}_\alpha$ planes, or shared $< a >_\alpha$ directions within each parent β grain. Analysis of the relationship between the α-phase domains within the parent β grain is similar to the analysis of parent austenite grain (PAG) block and packet analysis, as explored in (Morito *et al.*, 2003) (Morito *et al.*, 2003). Note that , the variant numbering for α-domains may be different within two adjacent parent β-grains.

#### 4. Simulated dataset

To validate the algorithm, a simulation was performed. A dataset consisting of 6 β grains was created with a 'checkerboard' morphology to test whether the algorithm would determine the exact parent β microstructure and whether the uncertainty was correctly evaluated. This data set is shown in Figure 5.

From this β-grain structure, an α microstructure was then created from this by assigning an α variant number to each of the 12 sub-grains contained within each parent grain.

To test varying the number of unique α variants present and the effect of shared planes or directions, the assigned α variant numbers were carefully selected for the simulations for each parent grain, as detailed at Table 2. The potential α variants for each parent grain orientation were then calculated and the appropriate orientation assigned to each α sub grain. The resultant α microstructure was then used as the input for the algorithm.

Table 2 - Simulated dataset α variant population choices and the number of potential parent β orientations that can be found based on these.

| **Grain No.** | **No of Unique α variants** | **Details** | **No of potential parent β orientations** |
|---|---|---|---|
| 1 | 12 | All 12 variants present | 1 |
| 2 | 3 | 2 common <a> + 1 random | 1 |
| 3 | 3 | All common <a> | 2 |
| 4 | 2 | No commonality | 1 |
| 5 | 2 | Both common (c) | 3 |
| 6 | 1 | Only 1 variant | 6 |

The simulated dataset and the corresponding reconstructed parent β microstructure are shown at Figure 5 (and are provided in the supplementary data). As can be seen in Figure 5 (C), the parent grain boundaries are correctly identified (red boundaries), as are all of the α-α boundaries within each grain (blue boundaries). The reconstructed parent grain microstructure is correct for each of the 'known' (1 potential β orientation) parent grains. However, as expected, there is uncertainty for the remaining grains. The chosen parent grain options can be selected to match the original parent β orientations, but the output may not select this (as there is a random selection of the represented β grain from the uncertain candidate orientations). For each uncertain grain, the alternative β orientation options are presented as the small circles within each grain in Figure 5 (D).

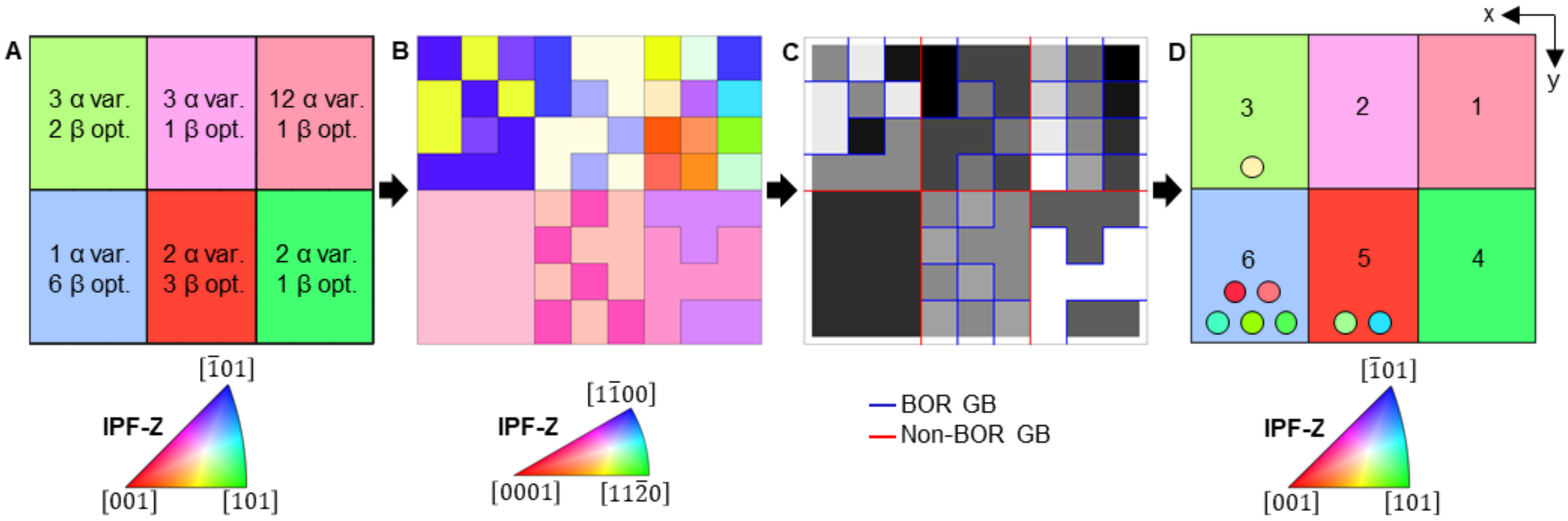

**Figure 5** Simulated dataset in (A) β phase (IPF-Z) and (B) α phase (IPF-Z). (C) Grain boundaries identified by type on a greyscale variant number background – red = parent grain boundary, blue = α/α grain boundary. (D) The reconstructed parent grain microstructure (IPF-Z) – where there are multiple β grain orientation options, the alternative orientations are shown in the circles overlaid on the grains. The grain numbers on (D) correspond with the those reported in Table 2.

### 5. Experimental materials & method

For the experimental example, a Zircaloy-4 sample with a grain size approx. ~11 µm was used (from the same batch of materials as used by Tong & Britton (2017). The sample was encapsulated in a quartz tube, backfilled with Ar, and heated to >1000 °C (i.e. super transus) for 10 mins and then water quenched. The sample was then ground using 800-2000 grit SiC papers followed by a 4.5 hr polish with OP-S. Finally, the sample was broad ion beam polished using a Gatan PECSII machine for 15 mins at 8 keV, 8° tilt, 1 rpm, dual modulation.

EBSD data was captured using a Quanta FEG 650 SEM equipped with the Bruker eFlashHD2. Data was processed using Bruker ESPRIT software and data exported to .h5 prior to analysis in MATLAB (R2018b) using MTEX version 5.4. EBSD patterns were captured using a 20 kV beam with a working distance of 15 mm and 70° sample tilt. The patterns were captured at a resolution of 320 x 240 pixels using a 34 ms exposure time and 0.14 µm step size.

## 6. Results

The reconstruction of this experimental data is shown in Figure 6, which includes the α orientations, grain boundary map with labelling of the BOR boundaries, and associated parent β microstructures. For the parent β microstructures, the initial reconstruction output (Figure 6 (3)) is based on the α grain groupings identified using the MCL algorithm - the grain boundaries present are the non-BOR boundaries identified at step 2. The reprocessed map (Figure 6 (4)) takes the orientations from the initial reconstruction map and processes the grains.

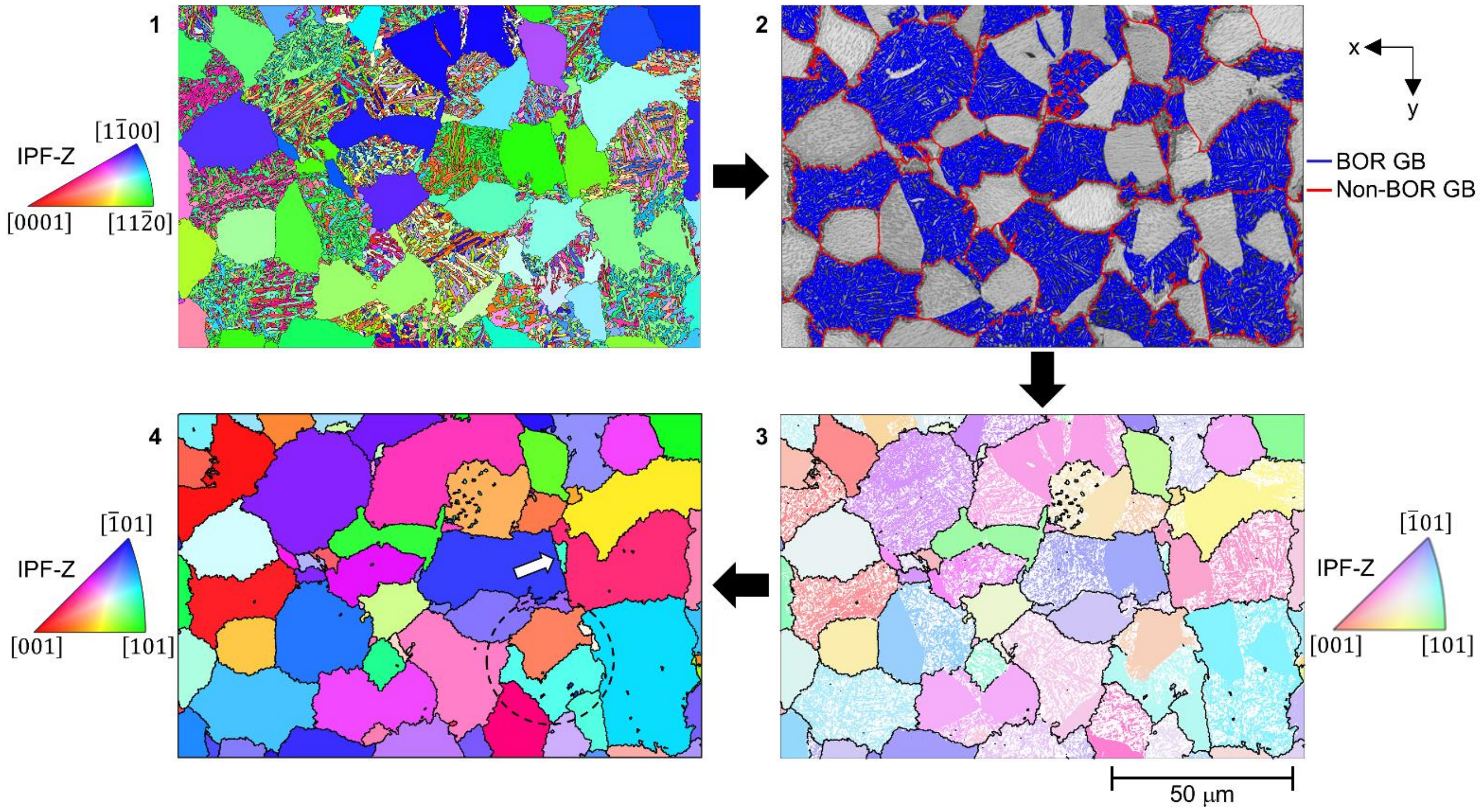

**Figure 6** Reconstruction of an EBSD map from a β quenched Zircaloy-4 sample showing [1] Initial α microstructure (IPF-Z) [2] Pattern quality map with overlaid grain boundaries, identified by type: RED = parent grain boundaries; Blue = α-α (or BOR) grain boundaries [3] Initial reconstruction output – β phase (IPF-Z ) [4] Reprocessed parent β grains (IPF-Z) with an arrow showing the location of the grain selected for further analysis. The grain pair within the dashed circle is explored in the Appendix.

This microstructure has (large) globular α domains, which have an approximate diameter of 20 μm. The origins of these globular domains is unclear. The microstructure we observe here is similar to many Ti-6Al-4V microstructures (Warwick *et al.*, 2013), where the global α-grains are thought to originate due to recrystallisation of α-laths in the α+β phase field. In our experiment, this globular α-nucleation and growth could occur as the sample is taken from the furnace and dropped into the quenching fluid. While the precise mechanism of this process is out of scope of the present work, it is interesting to note that often these globular grains have an orientation that is related to the neighbour α-lath region.

Between these globular regions, there is a fine scale secondary α lath microstructure with multiple different α variants. The size and shape of these regions containing secondary α regions is of similar size to the globular α, but the α laths tend to have a width of ~1 μm. Analysis of the grain boundary map shows that the globular α and variants of α within the transformed secondary α domains are often from the same parent β orientation, and this is shown within the reconstructed parent β orientation maps in Figure 6.

An artifact was revealed in this reconstruction, as highlighted by the dashed circle in Figure 6 stage 4. In the reconstruction shown, this is labelled as two parent β grains. However, in this reconstruction, the parent β grain boundary has a shared $\{110\}_\beta$ and this explains why adjustment of the inflation power and cut off tolerance angle can result in this region being reconstructed a single (larger) parent β grain. For more information on this analysis, please see the appendix.

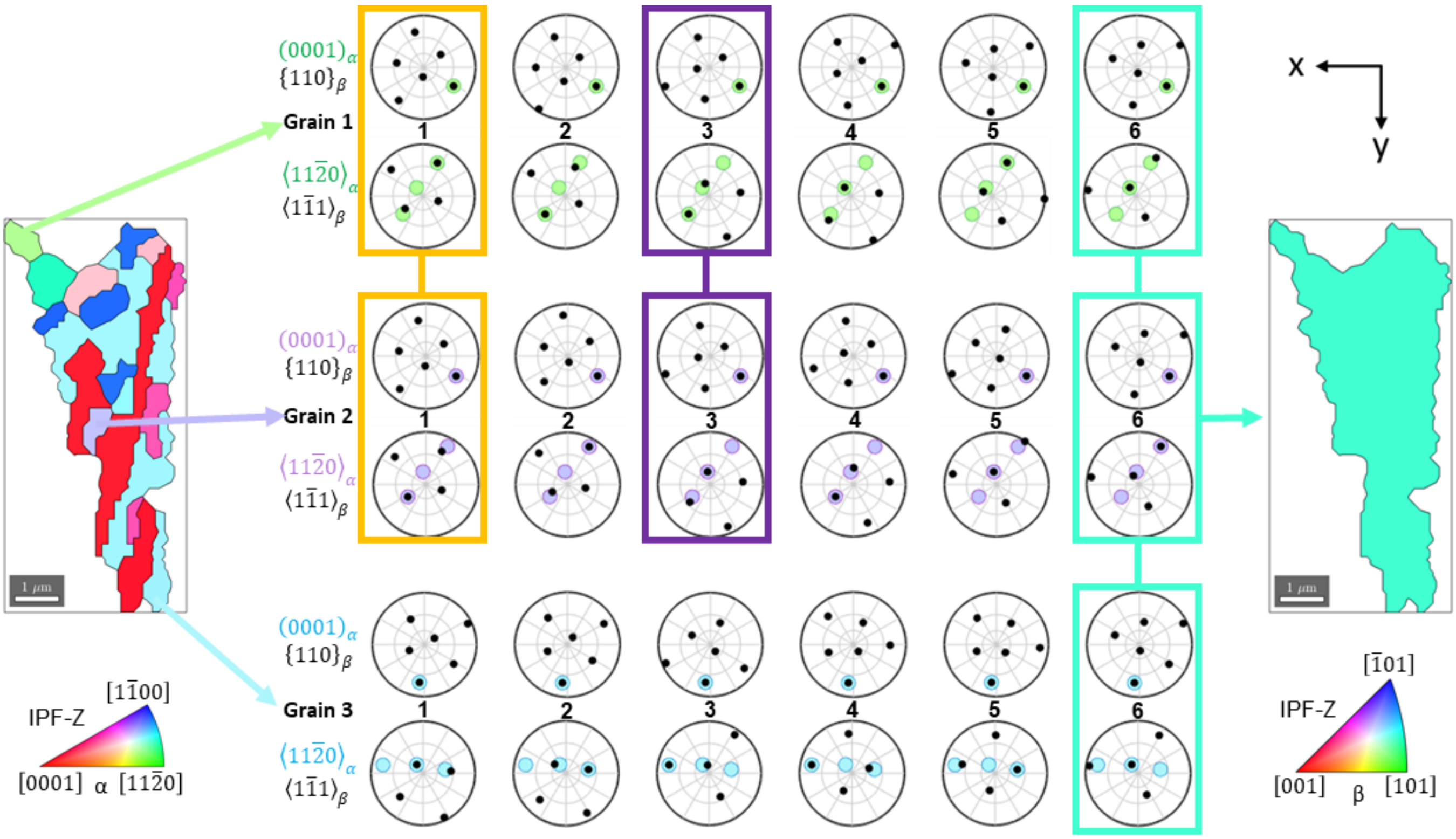

**Figure 7** Example showing how the β orientation certainty is affected by the number unique α variants present. There are three potential β variants (shown by the coloured & linked boxes) with just α grains 1 and 2, whereas with all three grains this reduces to one orientation (β variant 6 for all grains). α grain orientations are shown by the coloured dots in the pole figures, whilst the 6 potential β orientations for each of these are shown by the black dots in the pole figures.

To explore the certainty of the parent β reconstruction, one region was extracted from the map in Figure 6 (4) and the ORs are plotted in Figure 7. In this example, the α phase structure includes 8 variants of α phase, and so the parent β orientation is over determined. However, we can analyse the stereographic relationships between the α variants and parent β orientation to show which combinations of α variants which are required to determine a unique parent β orientation.

Three variants are highlighted, with the stereographic projections of the final (i.e. correct) β and α phase crystal planes and directions which are involved in the BOR. The combination of grain 1 (light green) and grain 2 (lilac) results in three potential parent β crystal orientations, as they have a shared basal plane. This ambiguity is immediately resolved when a third variant is identified from the same parent β grain if this (new) variant does not share the same basal plane.

Examination of the other 5 α orientations (not shown in figure) confirms that the β orientation determined is consistent with all the child α variants found in this region.

### 7. Post processing

The parent β reconstruction algorithm enables us to explore the relationship between the α grains within the complex microstructures. As we have demonstrated that the parent β microstructure can be determined using this algorithm, we can use it to explore the relationship between microstructural units within the α phase map.

In terms of mechanistic understanding, it is useful to report on four relationships:

(1) The number of potential parent β orientations, i.e. the uniqueness of the parent β reconstruction.
(2) A labelled map of each α variant within each β grain.
(3) Sharing of the basal plane within each parent β grain.
(4) Sharing of the <a> direction within each parent β grain.

For example, four maps for these relationships (using the experimental dataset) are shown in Figure 8. These maps are related to each other, and also can be insightful depending on the microstructural analyses required.

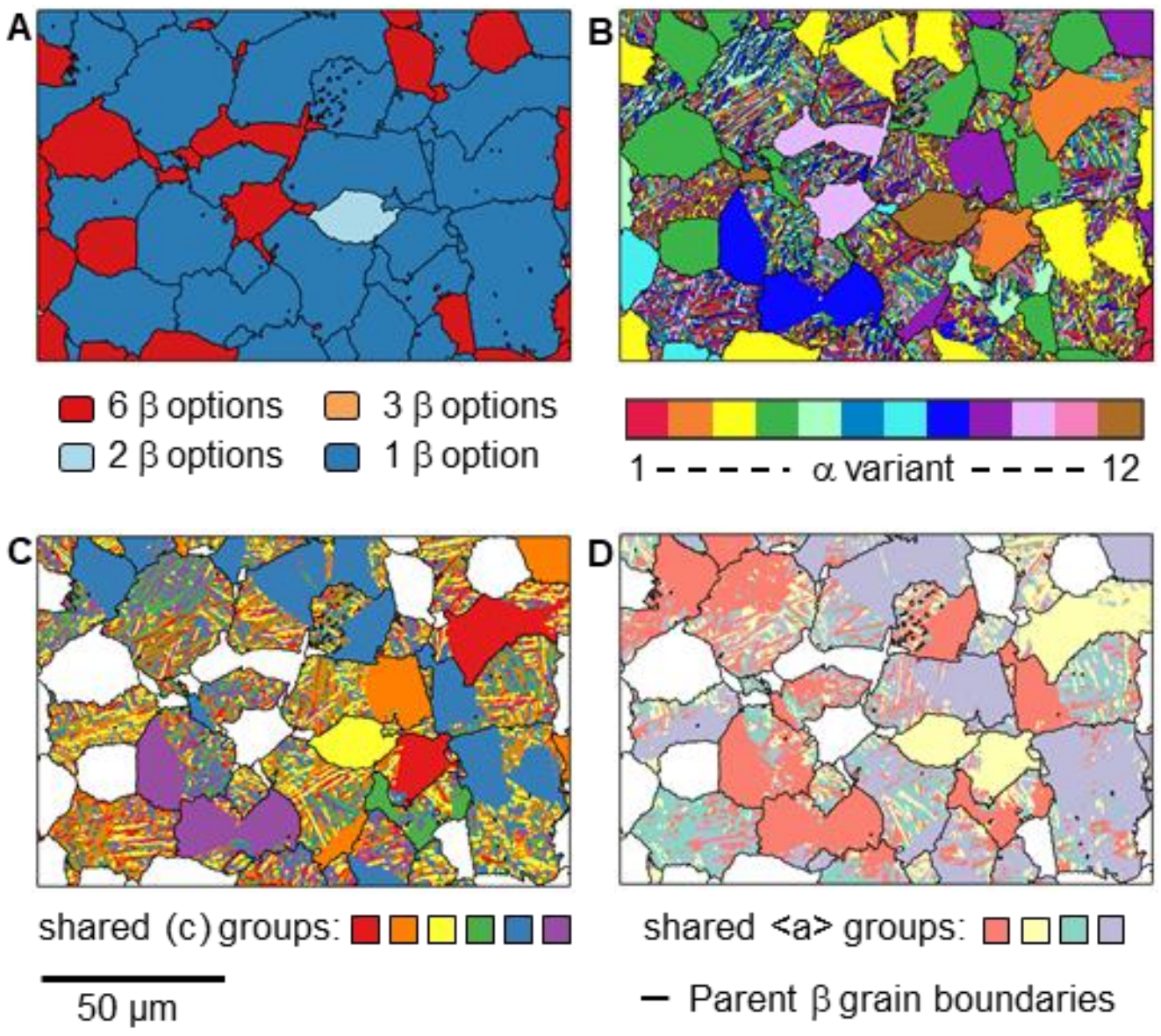

**Figure 8** Post processing outputs for β quenched Zircaloy-4 samples [A] β orientation certainty map showing the number of potential β variants for each parent β grain [B] α variant map – all 12 variants shown [C] α variant map coloured by α variants with shared (c) planes [D] α variant map coloured by α variants with shared <a> directions. Note: for [B]-[D] comparisons can only be made directly within the grains as it is calculated on a grain by grain basis.

Uncertainty of the parent β orientation is dependent on the number α variants generated from each parent β, combined with whether these variants share <a> directions and basal planes. This is shown pictorially in Figure 7. This ambiguity was also explored within the simulated data set shown in Figure 5. The uncertain cases, and the reason why they are uncertain (e.g. due to shared basal planes or shared <a> directions, can be inferred from the associated other figures in Figure 8, such as for the pale blue region (in Figure 8 (A)) where there are 2 two β options. In this case, there are two α variants present (the second grouped variant is a smaller grain not visible on the figure) which share a common <a>, hence the ambiguity.

The map of variants, shown in Figure 8 (B), can be useful to evaluate the size and shape distribution of variants within a parent β grain. For example, where the large globular α and lath α reconstruct to the same parent grain, the globular α variant type and the fine scaled transformed α can be compared (different or similar variants). The colour coding as presented plots the variants with respect to each parent β grain structure, as you cannot read the relationship between α variant numbering from one parent β grain to the next.

The map of shared basal planes is likely to be useful when considering properties that will extend across microstructural regions, such as any property that is transversely isotropic in the zirconium system (e.g. elastic stiffness). This map could also be useful to help navigate arguments in the literature on failure under fatigue loading and the formation of facets (Uta *et al.*, 2009). Again, similar to the variant analysis, this map must be interpreted with respect to the parent β grain structure, as a similarly coloured shared basal plane in this map across with two β grains is unlikely to have a common plane normal.

Finally, the map of shared <a> directions echoes the idea of the shared basal plane map, but now is true for the <a> directions. This map could be important when considering the propagation of slip through the microstructure and the relative transparency of different α/α grain boundaries. Furthermore, if there is a transformation strain associated with the β→α OR, potentially analysis of this map could assist in understanding transformation strain heterogeneity and variant selection (Shi *et al.*, 2016).

### 8. Comparison with MTEX

In the process of developing our algorithm, a reconstruction algorithm was developed within MTEX.

The MTEX reconstruction was performed using MTEX 5.7.0. The following recipe was used:

- calculate (α) grains (4° threshold used for direct comparison), remove quadruple points and remove small grains (less than 5 points);
- Compute parent orientations from triple junctions (& probabilities);
- Use a voting system (threshold 0.7) to determine which grains are converted to parent grains;
- Grow parent grains at grain boundaries;
- Merge parent grains;
- Merge inclusions.

A comparison of the reconstruction between the ParentBOR code and the MTEX reconstruction is shown in Figure 9. Of note, the MTEX reconstruction algorithm does not include any of the post processing analysis that we have described for the ParentBOR code. In practice, this could be included in the MTEX algorithm but it would need some subtle (but important) changes with respect to how the variants are numbered and the results are described within the MATLAB variable arrays.

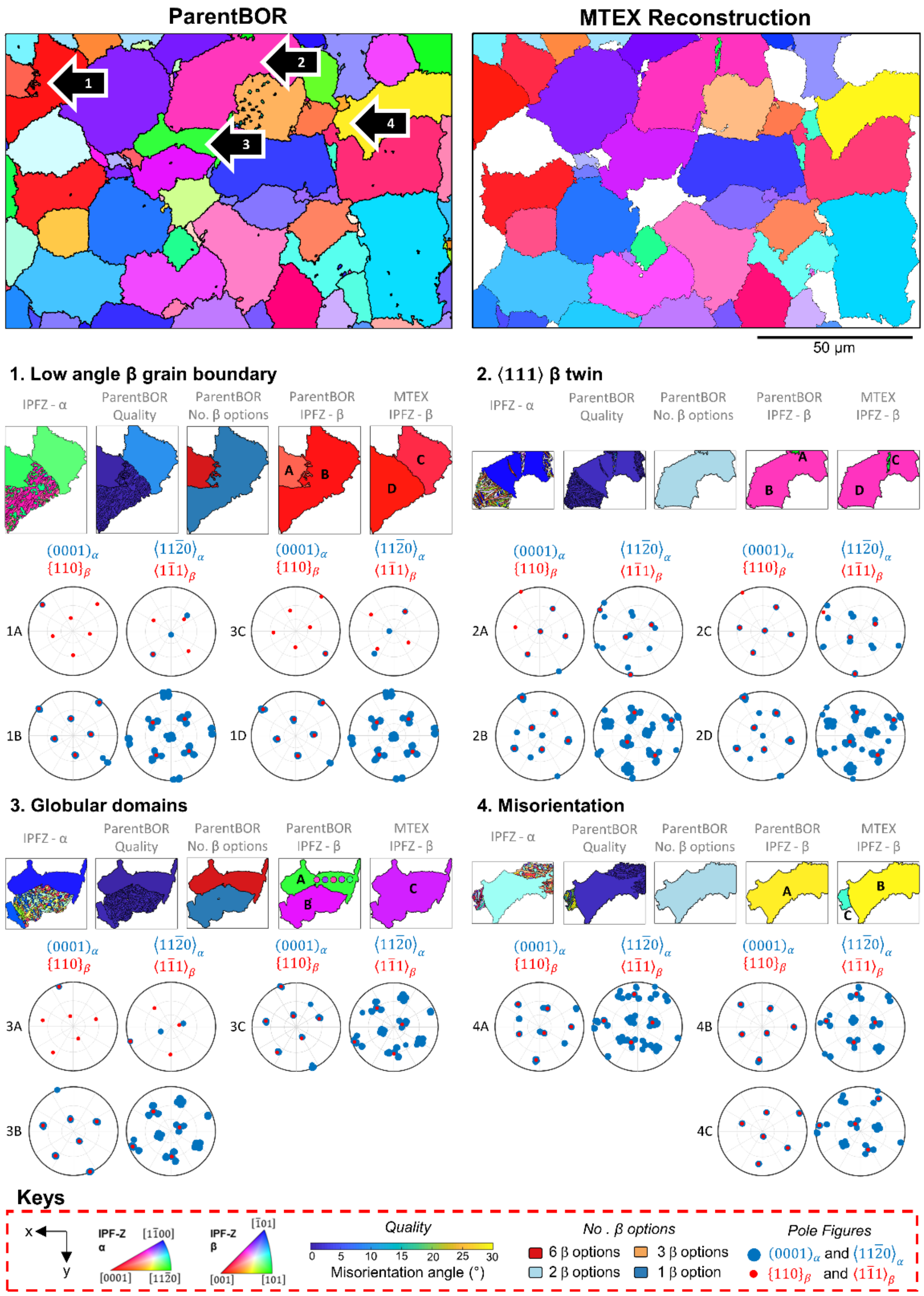

**Figure 9** Comparison of ParentBOR reconstruction and MTEX 5.7.0 reconstruction.

The MTEX reconstruction does not reconstruct regions where there are more than one parent β grain orientations, and these are shown as white regions in Figure 9. In addition to these (apparently unsolved) regions, there are four other types of difference between the two reconstructions:

### 8.1. Type 1 – low angle β grain boundary

The MTEX reconstruction and the ParentBOR reconstruction differ in how they build up the parent grains. In the MTEX reconstruction, variant orientations are considered with respect to triple junctions. In contrast, in the ParentBOR reconstruction domains are reconstructed using the MCL algorithm. In region 1, the large 'red' grain (i.e. near [001] out of plane) is broken down with two different sub-grain boundaries. In practice these regions are closely aligned, and they could be post processed to be joined together within the same large region of parent β structure.

### 8.2. **Type 2 - <111> β twin**

There is a small region within the pink grain which has a different orientation (in the parent reconstruction) to the neighbourhood. Analysis of the β pole figures reveals that this region is a <111> type twin in the parent β orientation. The orientation relationship between these two regions in the β phase, and the associated inheritance of this OR through the α domain, results in a slight confusion in the size and extent of the parent β grain size (and the very small grain within this parent β domain).

### 8.3. Type 3: Uncertainty of globular domains

The MCL clustering regime tends to grow domains outwards from clusters of similar orientation. Spatial regions which include a single variant can cause issues when the reconstruction is performed. Here we see that there is a region (A) which does not have a certain parent β orientation, but the candidate orientations of this region include both the green (the ascribed colour) as well as the purple orientation (closely aligned to region B, and similar in orientation to the MTEX domain, labelled C). This highlights that the ParentBOR code could be extended further to post process the parent β map, for regions that are less certain, to consider whether they belong to nearby regions.

### 8.4. Type 4: Misorientation

For this region, we observe that the ParentBOR reconstruction shows one parent β grain and the MTEX reconstruction reveals two separate parent β domains. Analysis of the pole figures shows that region A (the ParentBOR grain) has the same β orientation as region B (the large area from the MTEX reconstruction). In region A however, there is a region which has a high angular deviation for the left-hand shoulder (which MTEX describes as a different β grain). The poor reconstruction of this area is shown as a high value (i.e. bad) in the quality map. It is likely that the MCL algorithm clusters these two regions together despite there being a small misorientation between the α variants (and associated small angular deviation for the β grains).

## 9. Discussion

We present an evolved form of parent β-grain analysis, together with some new post processing capabilities.
The MCL algorithm to group the α grains parent to reconstruction provides a good balance of speed and accuracy, whilst the single coarsening parameter (inflation power) provides the opportunity to easily adjust the number of discrete clusters identified. This has benefits and disadvantages as there is the potential to add parent grain orientations if there are too many clusters identified. The inflation variable can average small sub-groups of grains within each parent, and this risks adding ambiguity to the analysis. This is also implemented within MTEX, and released as open source, which may promote wider use and further developments of the method.
In the reconstruction, we have selected to use all α grains present for our code, but this leads to uncertain parent grain orientations in cases where there are insufficient variants present. The post processor for the approach will subsequently label these regions and provide suggestions for the alternative β orientations which may be selected, as per the users' needs.
The simulated dataset (with code included in supplementary data) allows us to validate the reconstruction and test the post processing, and it can be adapted to different sized datasets. In development of our approach, the simulation has been useful to explore specific scenarios e.g. where there is uncertainty as to why a parent β grain reconstructs as two grains or an α grain flips between two parent β grains with changing inflation power. Application of the code to experimental data revealed further nuances, such as the ORs within the parent β structure that can cause confusion within the reconstruction.

We have also evaluated the ParentBOR code against recently available code within MTEX (see Section 8). While there are (understandable) differences between these reconstructions, it is also useful that we now have two analysis methods that are constructed within the same overall framework. This enables us to understand how reasonable (or unreasonable) the reconstruction is with regards to specific algorithm induced artifacts. The relevance of these artifacts will depend on how the parent β grain structure is interrogated. This comparison highlights that the ParentBOR code reconstructs a significant fraction of the map equally to the MTEX implementation, and that there are specific regions which are reasonably reconstructed in ParentBOR and vice versa. This implies that a hybrid approach could be implemented, such as the recently described methods of (Huang *et al.*, 2020) which include orientation relationship refinement, orientation coalescence and regional voting.

Recent advances using pattern matching by (Lenthe *et al.*, 2020) of overlapped BOR related patterns offer the chance to index finer scaled microstructures, where there may be two or more overlapping patterns within the interaction volume. Fortunately, our microstructure mapped sufficiently well (>83% success) and there is no retained β in this material.

From a materials engineering perspective, we expect that the post processing analysis is likely to prove useful for understanding variant selection during thermomechanical processing of alloys with a BCC to HCP allotropic phase transformation, such as titanium, zirconium, and hafnium. For the future, it may also be useful to explore the relative effectiveness of different grain boundaries for specific materials properties, such as the effective slip length in during deformation.

## 10. Conclusion

In this work, we have:

1. Introduced a HCP→ BCC parent phase reconstruction method and demonstrated this on simulated and experimental zirconium material data.
2. Created a post processor to reveal the uniqueness of the parent β grain orientations and shared crystallographic relationships within the parent β grains.

We found that the MCL method can be adapted to different ORs successfully, and the addition of a post processing step (specific to the materials system) can be useful to understand associated ambiguity of the parent microstructure. We hope that the post processing tools, which have been designed to explore shared basal planes and shared <a> directions, will be useful in understanding materials performance issues used in engineering applications.

## 11. Data availability

To further work in this field, we also provide our algorithms and data. The ParentBOR analysis code is available via https://github.com/ExpMicroMech/ParentBOR and the example data can be found via Zenodo (DOI:10.5281/zenodo.4632039).

## 12. Author contributions

RB conducted the experiments, developed the code, and wrote the first draft of the manuscript. TBB supervised the project.

**Acknowledgements** We acknowledge helpful discussions with Professor Fionn Dunne and Dr Katharina Marquart. TBB acknowledges funding from the Royal Academy of Engineering for his Research Fellowship. TBB and RB acknowledge funding from the EPSRC EP/S515085/1 and Rolls-Royce plc for providing material. We thank Dr Thibaut Dessolier, Dr Tianhong Gu, and Ms Sarah Hiew Sze Kei for proof reading and comments on the final draft.

**Appendix A.**

The parent β grain identified by the dashed circle in Figure 6 (4) can be reconstructed as a single parent β grain, or two (main) parent β grains. This uncertainty is caused by a closely aligned $\{110\}_\beta$ plane between the two parent β grains as shown in this case. Whether it reconstructs as one or two grains depends on the number of discrete clusters formed prior to reconstruction, in other words, the inflation power. This is shown in Figure 9 where a smaller region from the larger map has been selected and reconstructed using a range of inflation powers.

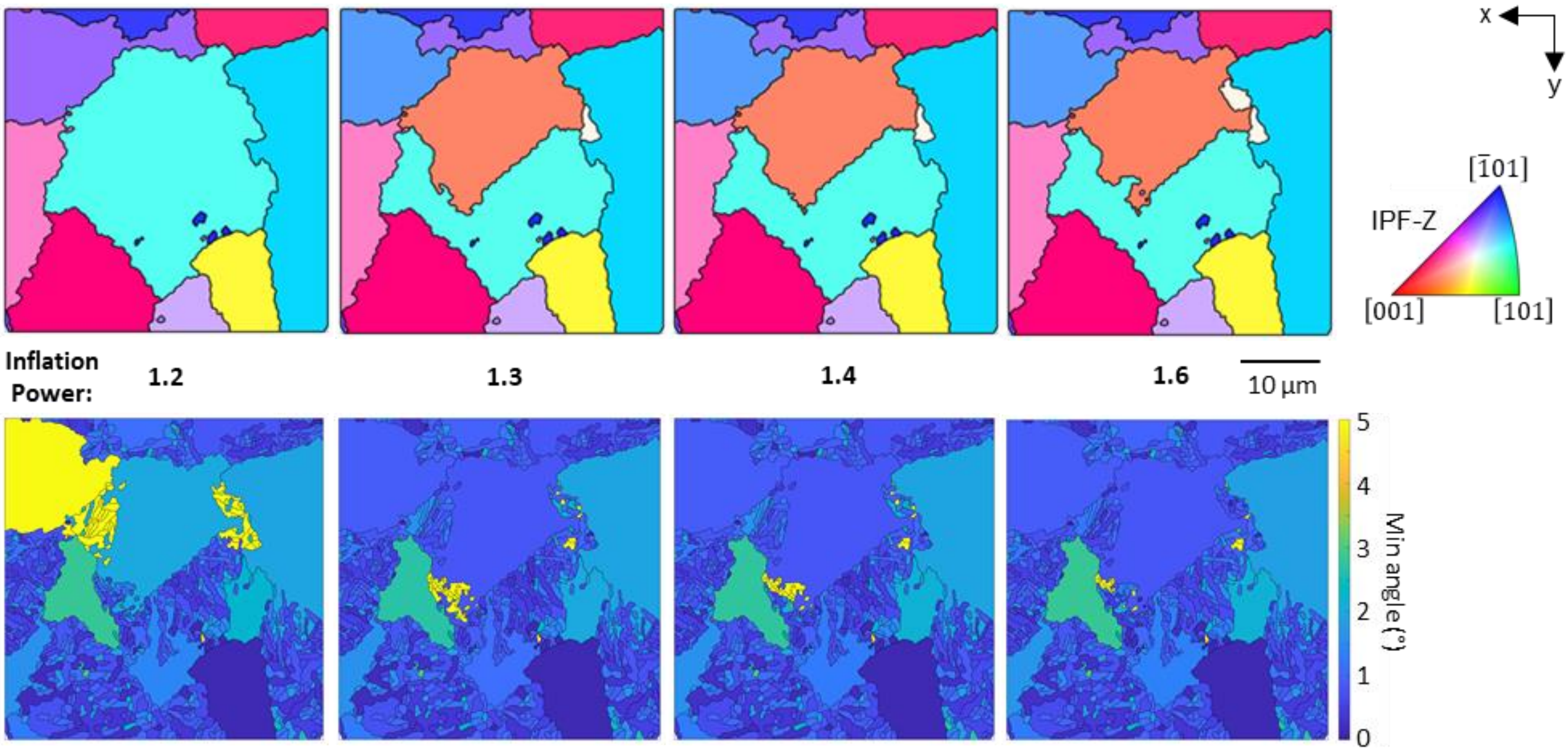

**Figure 10** Reconstruction of smaller region into one or two (main) parent β grains - effect with changing inflation power from 1.2 to 1.6. In this figure, the upper row shows the reconstructions as coloured using the IPF-Z colourmap, and the bottom row shows the calculated misorientation of the (as measured) α grains from the ideal α orientations determined from the parent β grain orientation.

Looking at the two β grain option, the two orientations have a shared $\{110\}_\beta$ plane, with a 1.34° angle between the two $\{110\}_\beta$ planes and a total misorientation of 47° between the two crystals (i.e. they are rotated about the shared $< 110 >_\beta$). This is shown visually in Figure 10 (A) where the two parent grain orientations have been plotted as overlapping cubes, with the common planes indicated by the solid colours. The consequence of this shared plane is that there are alpha variants which would match either parent grain orientation – this is where the ambiguity in the reconstruction occurs.

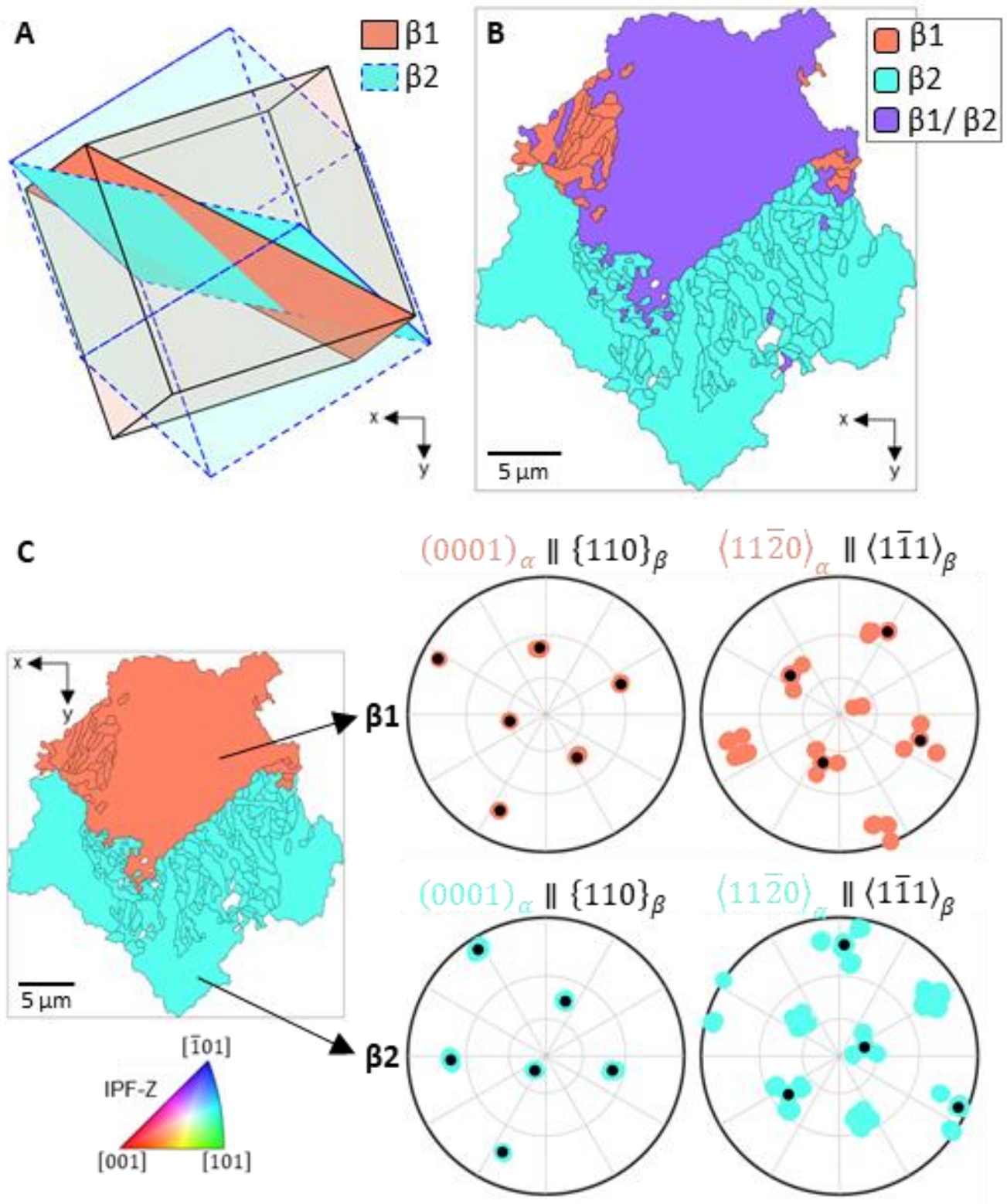

**Figure 11** (A) Comparison of the two β grain orientations, showing the shared planes for each using solid colour (corresponding to the IPFZ map colours for each grain). (B) alpha grains for this region coloured by parent grain orientation β1, β2 or uncertain (β1/β2). (C) manually sorted α grains, coloured by parent β grain assigned (IPFZ colouring) – along with the pole figures for each parent β grain region, with the alpha orientations the coloured dots and the α orientations the black dots in each case.

A manual analysis of the αgrains contained within this region, in Figure 10 (B), shows the issue more clearly, with uncertain grains shown in purple. The region could plausibly be reconstructed as a single parent grain, although there would be regions (orange grains in Figure 10 (B)) that don't match this parent grain well – these could potentially be other parent β grains. However, a better reconstruction (comparing the alpha and β orientations on pole figures) can be achieved using the split grain option, as demonstrated by Figure 10 (C). It is therefore likely that the 2 parent β grains, with a shared $\{110\}_\beta$plane is a valid reconstruction.